\documentclass[a4paper]{jpconf}
\usepackage{graphicx}
\usepackage{subcaption}
\usepackage[utf8]{inputenc}
\usepackage{hyperref}
\begin{document}

\title{Optimization of a fast rotating target to produce kHz X-ray pulses from laser-plasma interaction}
\author{L Mart\'in\textsuperscript{1}, J Benlliure\textsuperscript{1}, D Cortina\textsuperscript{1}, J J Llerena\textsuperscript{1}, D Gonz\'alez\textsuperscript{1} and C Ruiz\textsuperscript{2}}

\address{\textsuperscript{1} IGFAE, Universidade de Santiago de Compostela, 15782 Santiago de Compostela, Spain.\\
\textsuperscript{2} Instituto Universitario de F\'isica Fundamental y Matem\'aticas y Departamento de Did\'actica de la Matem\'atica y de las Ciencias Experimentales, Universidad de Salamanca, Patio de Escuelas s/n, Salamanca, Spain.}

\ead{lucia.martin@usc.es}

\begin{abstract}
We report the development of a fast rotating target to produce ultrashort incoherent X-ray pulses from bremsstrahlung. These short X-ray pulses are produced in the laser-plasma interaction of a 35 fs, 1 mJ pulse of a Ti:Sa laser of 1 kHz repetition rate, with a solid metallic target. In this paper, we report our developments to improve the stability of this micron size source of ultrashort X-rays. As the Rayleigh length is very small (\textless15 $\mu m$), wobbling of the rotatory stage can reduce the intensity on target and change the characteristics of the source. We describe the methods we have developed to measure and adjust the stability of the focus on target. These advances are important for the development of sources with high average power and good stability. The X-ray source has a broad Maxwellian-like distribution with temperatures of around 10 - 40 KeV and could be used for advanced X-ray imaging such as absorption or phase contrast tomography.
\end{abstract}

\section{Introduction}
Femtosecond hard X-ray sources are an important tool to describe the ultrafast dynamics in physics \cite{wachulak,adeline}, chemistry \cite{legall}, biology \cite{wenz} and material science \cite{Silies:2009gw}. Optical pump and X-ray probe have demonstrated to be an important test for the sensitivity of detection schemes and provide structural and dynamical information of the systems under study.\\
Laser-driven table-top X-ray sources have been demonstrated in solid \cite{gobet,bixue,li} and gas media \cite{wenz,zhangl}, and a new family of all optical sources are being developed with high power lasers \cite{zamponi2}. Also, compact X-ray sources based on the $\lambda^3$ regime  \cite{naumova,naumova2} have been demonstrated in a wide variety of solid targets. These sources operate with a modest energy of few mJ and can reach KHz repetition rates which increase the average power of the source \cite{jannick}.

These X-ray sources are based on the interaction of an ultrashort laser pulse with an over-dense plasma. Combination of laser field and ionization during the first cycles of the laser leads to the creation of a plasma on the target surface \cite{Gibbon}. The density of this plasma is several times the critical density, containing hot electrons with  relativistic velocities due to the high intensity of the pulse ($I>10^{16}$ W/cm$^2$). These electrons interact with the heavy ions of the target as they leave it, producing Bremsstrahlung and characteristic radiation \cite{mccall,salzmann,galy}. These collisions lead to the heating of the target, increasing the plasma pressure and producing the ablation of the target surface \cite{Gibbon}. 
\clearpage
As the laser drives the electrons only for a short period of time, the result is a well defined beam of X-ray pulses with a broad energy distribution and a short duration of hundreds of femtoseconds \cite{kieffer}.\\
In this paper, we focus on the development of a fast rotating target which can sustain a stable and robust X-ray source in the $\lambda^3$ regime \cite{hou,bixue2}. The final objective is to increase the average power of the source and to obtain long operation periods supporting real world applications that would benefit special characteristics of this type of X-ray source \cite{jiang}. The mechanical stability of the target and the implementation techniques to diagnose the position of the target with respect to the laser focus, are the key issues to produce a X-ray source with these characteristics.

\section{Experimental set up and target positioning system for the kHz X-ray source}

\begin{figure}
\begin{center}
\includegraphics[width=0.92\textwidth]{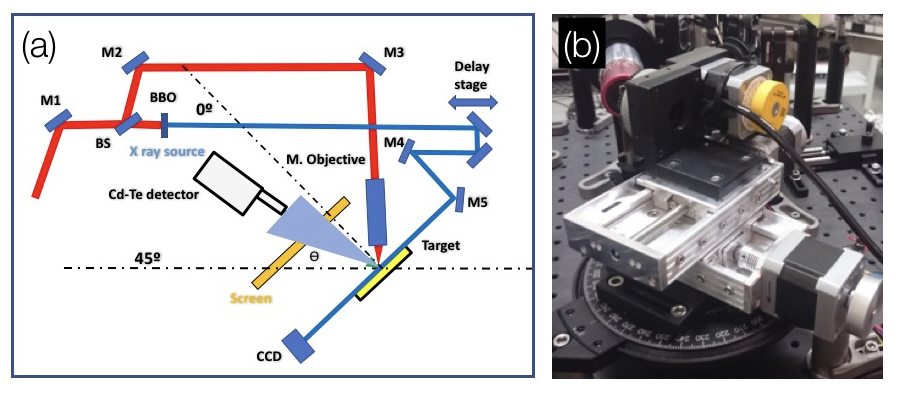}
\end{center}
\caption{ (a) Diagram of the experimental setup.  (b) Picture of the linear and rotating stages of the target system.} 
\label{expsetup}
\end{figure}

A back-scattered X-ray laser driven source has been recently installed at the Laser Laboratory for Acceleration and Applications (L2A2) at the University of Santiago of Compostela. We report the production of X-ray pulses at 1 KHz repetition rate from the relativistic interaction of an intense laser pulse and a solid target in air. The target is a flat copper metallic plate (55 mm x 55 mm) of 1 mm  thickness with a mirror like surface where the laser pulse impacts. In order to have a stable X-ray source, the target needs to accomplish two conditions simultaneously: The positioning system should refresh the target material, because the material is destroyed after each shot. During the interaction, a large number of electrons are expelled from the material leading to the destruction of the target of an area  much larger than the laser focus. Therefore, the target positioning system should rapidly replace the material to sustain a KHz repetition rate. The positioning system should also keep the target at the focal plane of the laser as the material is refreshed. Because we use a tight focusing scheme to obtain the smallest possible focal spot, the Rayleigh length is small (\textless15 $\mu m$) and the tolerance in positioning must be kept below the Rayleigh length.\\
\begin{figure}
\begin{center}
\includegraphics[width=.9\textwidth]{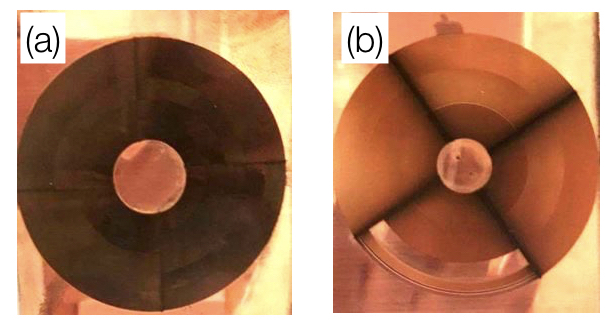}
\end{center}
\caption{Two targets irradiated in different conditions. (a) Target irradiated with constant angular velocity  and (b) variable angular velocity.}
\label{targets}
\end{figure}

The experimental setup is shown in Fig.\ref{targets}. An ultrashort laser pulse (THALES ALPHA 10/XS, 1 mJ, 35fs, 1KHz, 800 nm) is first divided through a beam-splitter (BS, 90R:10T) and then transported with M1, M2 and M3 dielectric mirrors (Thorlabs, BB1-E03) towards the microscope objective (Mitutoyo, M Plan Apo NIR 20X) with a focal length of 10 mm and Rayleigh length \textless 15 $\mu m$. \\

The laser pulse is focused on target at $45^\circ$ to produce a back-scattered X-ray beam close to the normal direction to the target. The diameter of the focal spot is estimated to be 3.5 $\mu m$, but a 5 $\mu m$ defocussing results in bigger spots up to 3.8 $\mu m$. In addition to this, the incidence of the laser beam is at $45^\circ$ so the effective spot diameter is between 4.9 and 5.4 $\mu m$. These spot sizes result in intensities on target between 1.1x$10\time 10^{17}$ W/cm$^2$ and 9.4x$10\time 10^{16}$ W/cm$^2$. The X-ray beam is analyzed with a Cd-Te spectrometer (Amptek, XR-100T-CdTe) to measure the X-ray energy spectrum. We use a 0.3 mm aluminum attenuator and a 1.15 mm lead collimator to minimize pile-up effects and to avoid the laser beam to entering detector. 
\\

The target positioning system and the diagnostic of the position are two of the most critical elements of the setup. The performance and stability of the X-ray source depends on the material refreshing and positioning of the target on focus during source operation. There are several solutions that have been used for this problem \cite{adeline,legall,li,zamponi2,zhangl}. We have chosen a positioning and refreshment system that consists on two linear stages (PiMicos VT-80) to change the longitudinal and perpendicular focal position on target, and one rotation stage (PiMicos DT-50). The target is attached to the rotating stage which is perpendicular to the optical board. The rotating stage is also attached through an L-shape piece to the two linear stages as shown in the Fig.\ref{expsetup}. For the measurements presented in this paper, we use laser pulses with an energy of $E=800$ $\mu J$ focused on target at $45^\circ$ incidence. The target material is fixed to the rotatory stage with double side tape. In this way, we minimize wobbling in the external radii of the pattern made by the movement of the target.\\

As the laser impacts on target it damages a region whose size depends on the energy and angle of incidence of the laser, being in any case larger than the laser spot size on target. As the rotation stage moves, the impacts on the target draw a circle as the stage complete a turn. Then, the first linear stage moves the target perpendicularly to the focal plane, in such a way that the next round of impacts draws a new concentric circle. \\

The second linear stage moves the target in and out of focus and it is used only to place the target in focus. This rotating target solution has important advantages compared to a Cartesian arrangement which is frequently used. In the case of a rotating target, there are no big accelerations in the movement, which favor the stability of the source. Moreover, linear stages are not as fast as rotating ones, so the setup presented here allows to minimize superposition of laser impacts at the target surface at 1kHz repetition rate, optimizing the X-ray dose.
The total number of shots per target depends on the distance between impacts and the available area on target. To maximize the number of shots per target and optimize the performance of the X-ray source we have established two strategies: The first strategy keeps the angular velocity of the target constant ($\omega$=$15^\circ$/s), which means that the linear velocity depends on the radius of the impact from the rotation center ($v=r\omega$). With this strategy the distance between impacts decrease with the radius.
In the second strategy, we keep the linear velocity constant ($v=196$ mm/s). As the perpendicular motor moves the target towards smaller radius, we change the angular velocity $\omega(r)=v/r$ using velocities between $450^\circ$$s^{-1}$ and $900^\circ$$s^{-1}$ to keep the separation between impacts constant.\\

\begin{figure}
\begin{center}
\includegraphics[width=.8\textwidth]{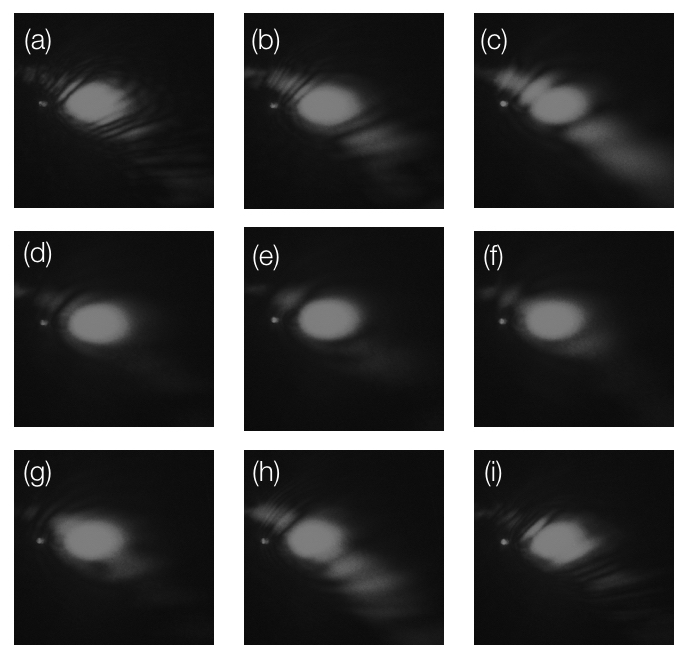} 
\end{center}
\caption{Speckle at different target positions. (a) 100 $\mu m$ after focus, (b) 50 $\mu m$ after focus, (c) 20 $\mu m$ after focus, 
(d) 5 $\mu m$ after focus, (e) Speckle at focus, (f) 5 $\mu m$ before focus, (g) 20 $\mu m$ before focus, (h) 50 $\mu m$ before focus and (i) 100$\mu m$ before focus.}
\label{speckle}
\end{figure}

Figure \ref{targets} shows two copper plates after being used as targets. Fig.\ref{targets}(a) corresponds to an irradiation following the first strategy of constant angular velocity, and Fig.\ref{targets}(b) shows the case of constant linear velocity. In both cases, the uniformity in colors shows that the source is always in focus producing a similar damage per shot. Irradiation at constant linear velocity requires the measurement of the initial radius where the laser impacts. We use two laser impacts on target separated by a $180^\circ$ rotation to determine the center of the rotatory stage. Operating with a constant angular velocity is simpler, because it is not necessary to know the radius. However, the X-ray production in this mode changes from one radius to another, because the linear velocity varies, and also the superposition of laser impacts on target surface increases as the radius decreases. In order to reach a stable operation in this mode, it would be necessary to use larger targets to avoid overlap between impacts. We discuss the effect on the spectra of the two operation modes in Section 4.

\section{Diagnostics for target positioning}
\begin{figure}
\begin{center}
  \includegraphics[width=0.7\textwidth]{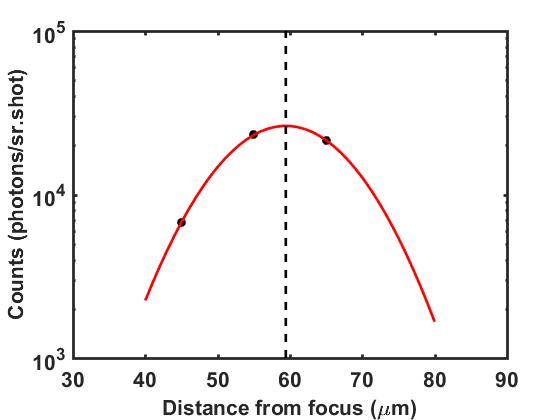}
\end{center}  
  \caption{Displacement of the nominal focal position due to self-focusing. The total counts for three spectra measured at different target positions are shown and used to calculate the best position (59 $\mu m$). }
  \label{selffocus}
\end{figure}

A stable X-ray source for this high repetition rate requires the minimization of the relative motion of the target surface  with respect to the laser focus as the target rotates. The Rayleigh length is small (\textless 15 $\mu m$), so we need to position the target with an accuracy better than 5 $\mu m$ to have the same intensity on target for each shot.\\

We use a speckle technique, to optimize the relative position of the target surface and the laser focus. For this purpose, we use a visible He-Ne laser (633nm, 1.2 mW, continuum) following the same path as the main infrared beam, and focused on target by the microscope objective. The specular reflection of the visible laser (speckle) is an homogeneous spot if the target is on  focus (Fig.\ref{speckle}(e)), because a 5 $\mu m$ diameter spot is smaller than the roughness of  target surface. \\

However, if the target is not on focus, the laser spot on the target surface will be larger and we will see the structure of the surface on to the speckle, becoming inhomogeneous (Figure \ref{speckle}). To position the target in focus, we observe the image of the He-Ne speckle on a white screen with a CMOS camera (5 Mpx, Mightex) while moving the target position with the longitudinal motor. The best focal position is determined when the cleanest speckle is found.\\

However, the implementation of this technique in air is more complicated, because the compressed infrared laser pulse (leaving the objective) suffers a nonlinear interaction with air before reaching the target surface. This nonlinear interaction makes the infrared laser beam converge before the visible one, resulting in the self-focusing of the infrared pulse due to the Kerr effect. In order to determine the best focal position, several spectra are taken at different distances from the He-Ne best speckle position.\\
We define the best focus position as the distance that maximises the total number of photons in the spectrum, while taking into account the self-focusing. This effect depends on the intensity, so for intensities around $10\time 10^{17}$ W/cm$^2$, the required displacement from the speckle position is about 59 $\mu m$ towards the objective as shown in Figure.\ref{selffocus}.

\section{X-ray source performance}
As a result of the laser-plasma interaction, a well defined X-ray conic beam is emitted from the target in a direction close to the target normal. The technique described in Sect.3 for the target-focus positioning has been used to investigate source stability at different radial positions. The X-ray spectrum is measured with a Cd-Te spectrometer at $15^\circ$-$17^\circ$ from the target normal.\\

\begin{figure}
\begin{center}
  \includegraphics[width=0.95\textwidth]{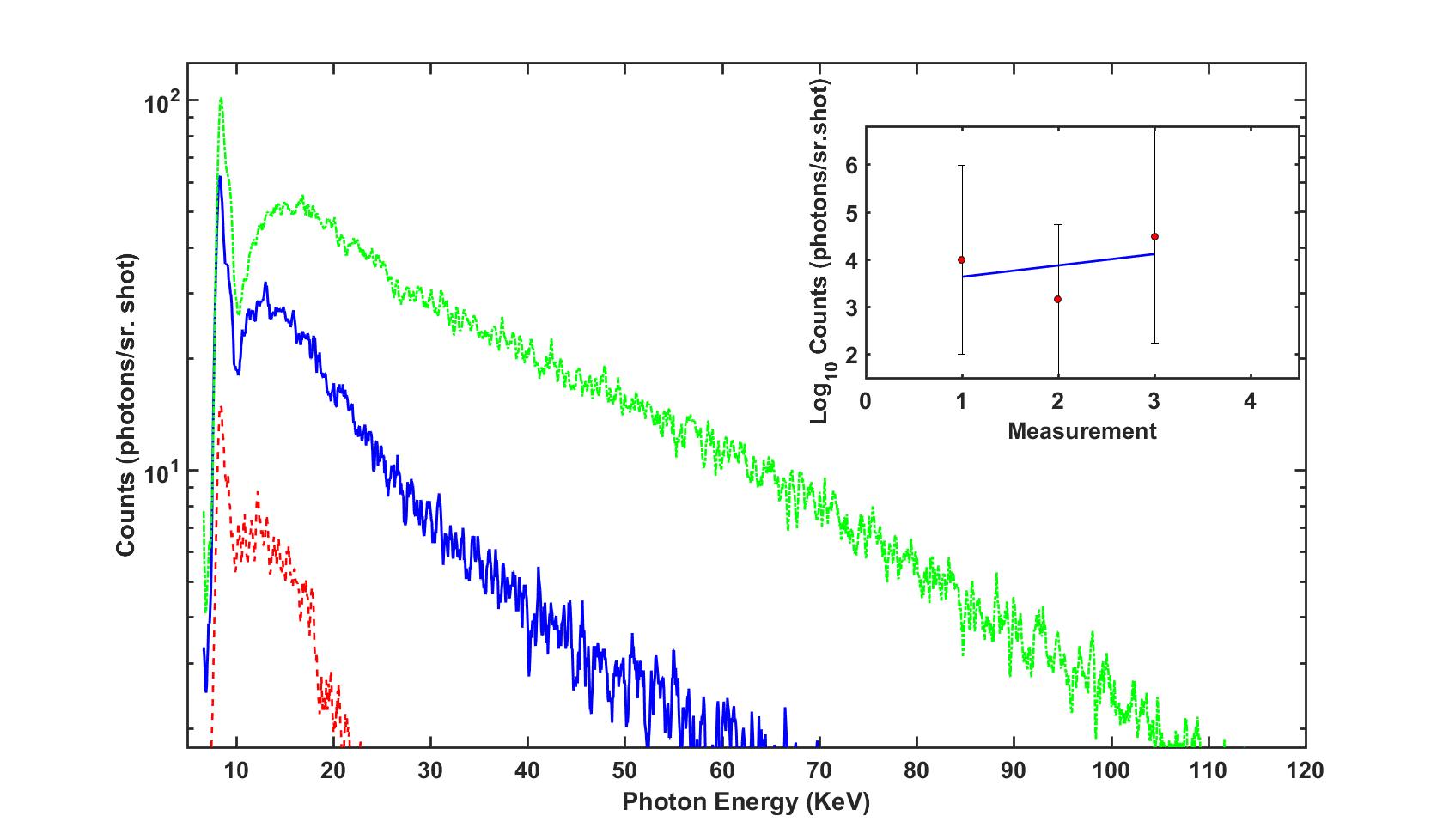}
\end{center}
  \caption{Three energy spectra for different radii and $I_{laser}$=1.2x$10\time 10^{17}$ W/cm$^2$,Cd-Te distance 24.5 cm and angle $15^\circ$ in constant angular velocity operation mode. (Inset image) Variation of spectra total counts 55\%.}
  \label{spectrum1}
\end{figure}

\begin{figure}
\begin{center}
  \includegraphics[width=0.9\textwidth]{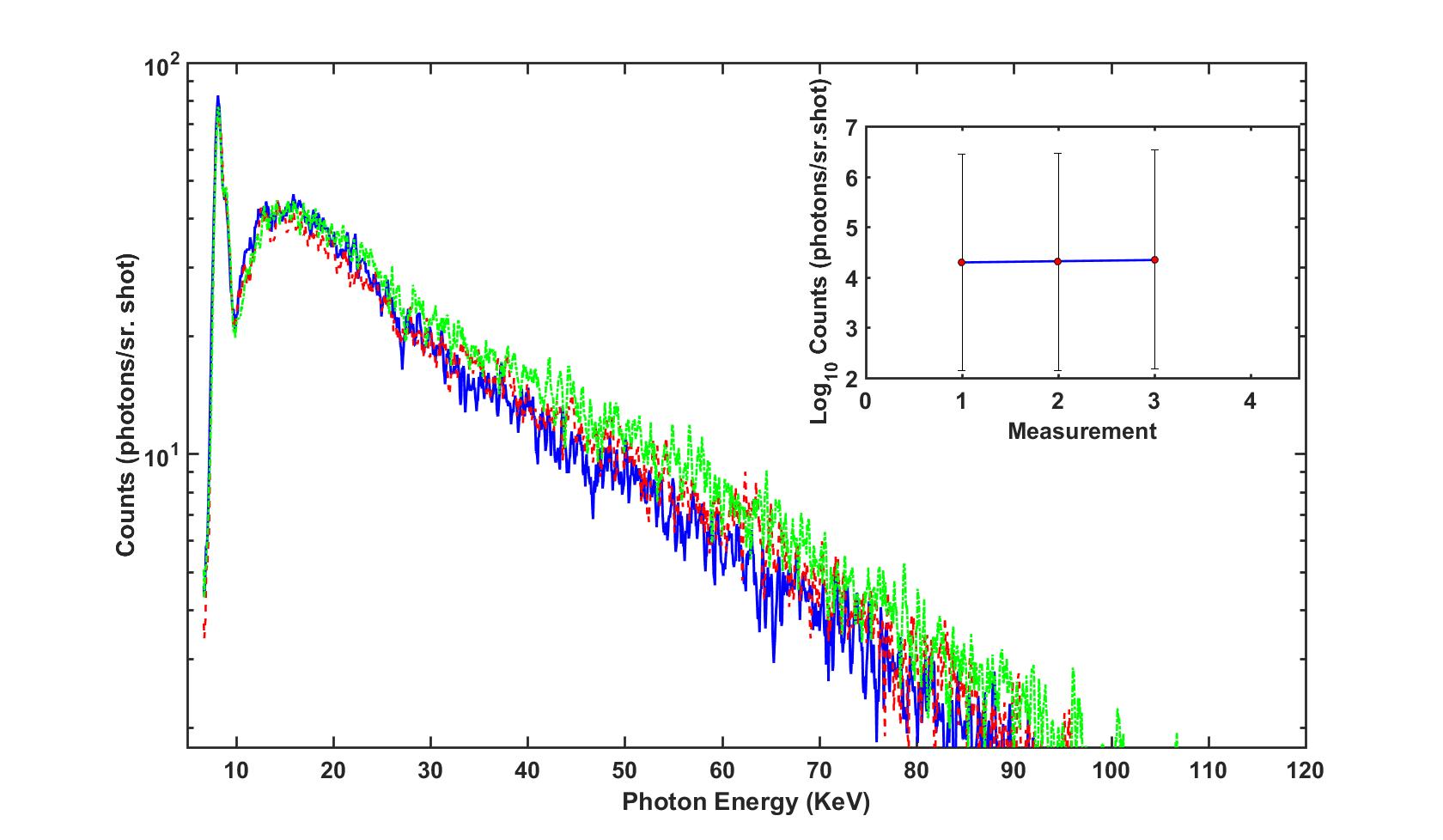}
\end{center}
  \caption{Three energy spectra showing the stability of the source taken at different radii and $I_{laser}$=1.1x$10\time 10^{17}$ W/cm$^2$,Cd-Te distance 24.5 cm and angle $17^\circ$) and variable angular velocity operation mode. (Inset image) Variation of spectra total counts.}
  \label{spectrum}
\end{figure}

\begin{figure}
\begin{center}
  \includegraphics[width=0.9\textwidth]{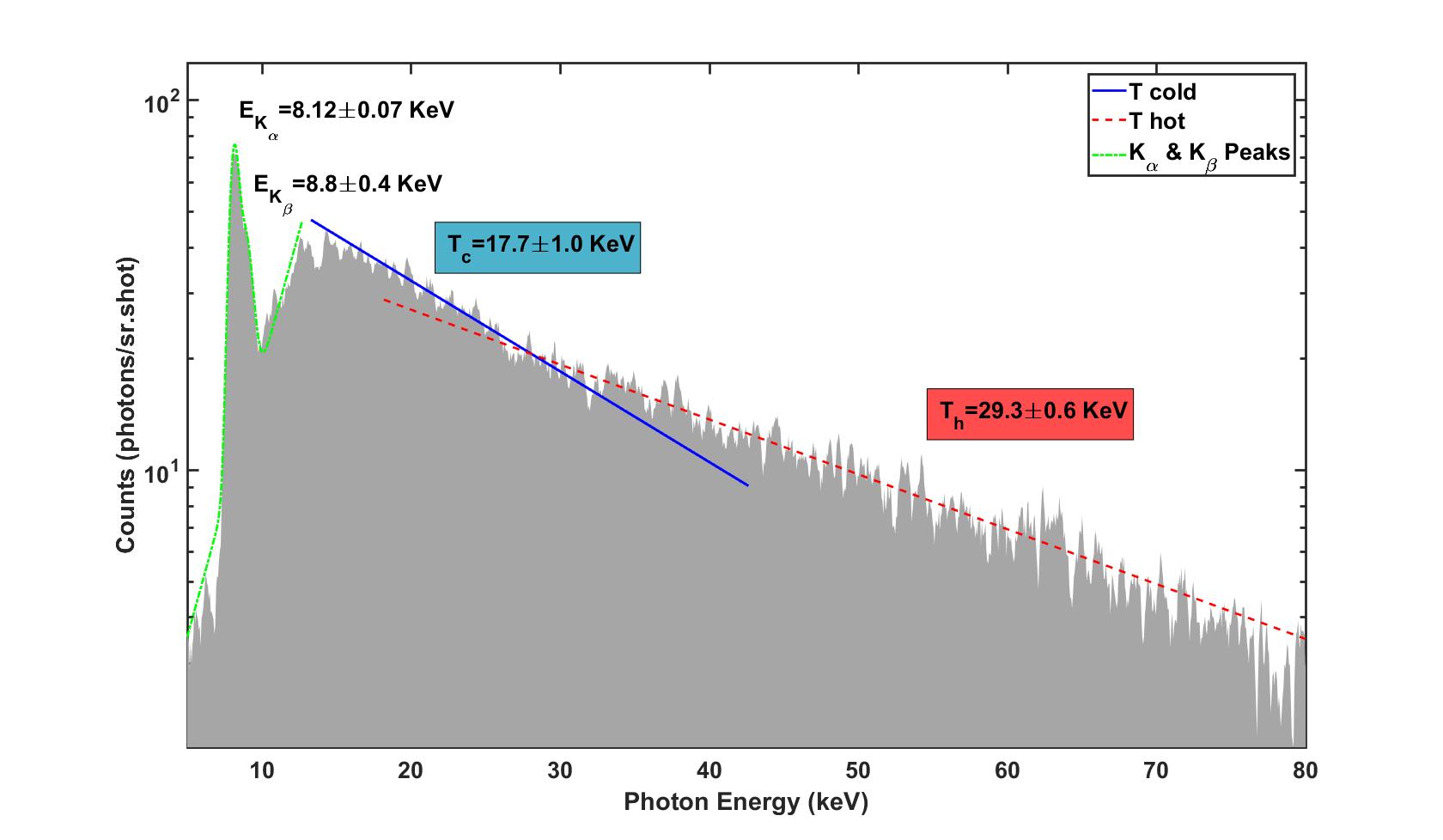}
\end{center}  
  \caption{Typical energy spectrum obtained irradiating a Cu target. In the spectrum we can clearly observe the Cu K peaks and also the cold and hot temperatures of the X-Ray continuum.}
  \label{temperature}
\end{figure}
\newpage
We can see the effect on spectra obtained with the same conditions and constant angular velocity in Figure \ref{spectrum1},  where deviation in the total number of between the measurements at different radii is above 50\%. However, if we adjust the rotation speed to the radial position of the laser impact on target, we obtain a very stable X-ray source. Indeed, Figure \ref{spectrum} illustrates that the reproducibility of the X-ray energy spectra in this case is better than 6\%.\\
In Figure \ref{temperature}, we report a typical X-ray energy spectrum, obtained with a Cd-Te spectrometer at $17^\circ$ from target and a laser intensity of 1.1x$10\time 10^{17}$ W/cm$^2$. The observed spectrum presents two peaks of copper characteristic radiation, $K_{\alpha}$ and $K_{\beta}$ and a Bremsstrahlung continuum with a bi-Maxwellian like distribution. This distributions consists on two temperatures $T_{cold}$ and $T_{hot}$ which are related with the different dynamics of the electrons inside the over-dense plasma. We obtain 8.12 and 8.8 keV for Cu K peaks, and $T_{cold}$=17.7 keV and $T_{hot}$=29.3 keV.

\section{Foreseen applications}
These kind of novel X-Ray sources driven by laser-plasma interaction open new possibilities of high quality X-Rays at small facilities. The principal advantages of laser-plasma X-ray sources are the smaller size of the source, and the short duration of the X-ray pulses. The short duration of the X-ray pulses (hundreds of femtoseconds to picoseconds) makes possible to perform X-ray time resolved images, which is interesting in a large number of research fields. Compared with conventional X-Ray tubes used for imaging, the smaller source size increases the spatial coherence of the emitted X-rays, making possible to implement more sophisticated imaging techniques than absorption imaging such as phase contrast imaging. Indeed, the quality of conventional absorption images is also improved, because the increased spatial coherence decrease the image blurring, resulting in better image quality.
\begin{figure}
\begin{center}
  \includegraphics[width=0.45\textwidth]{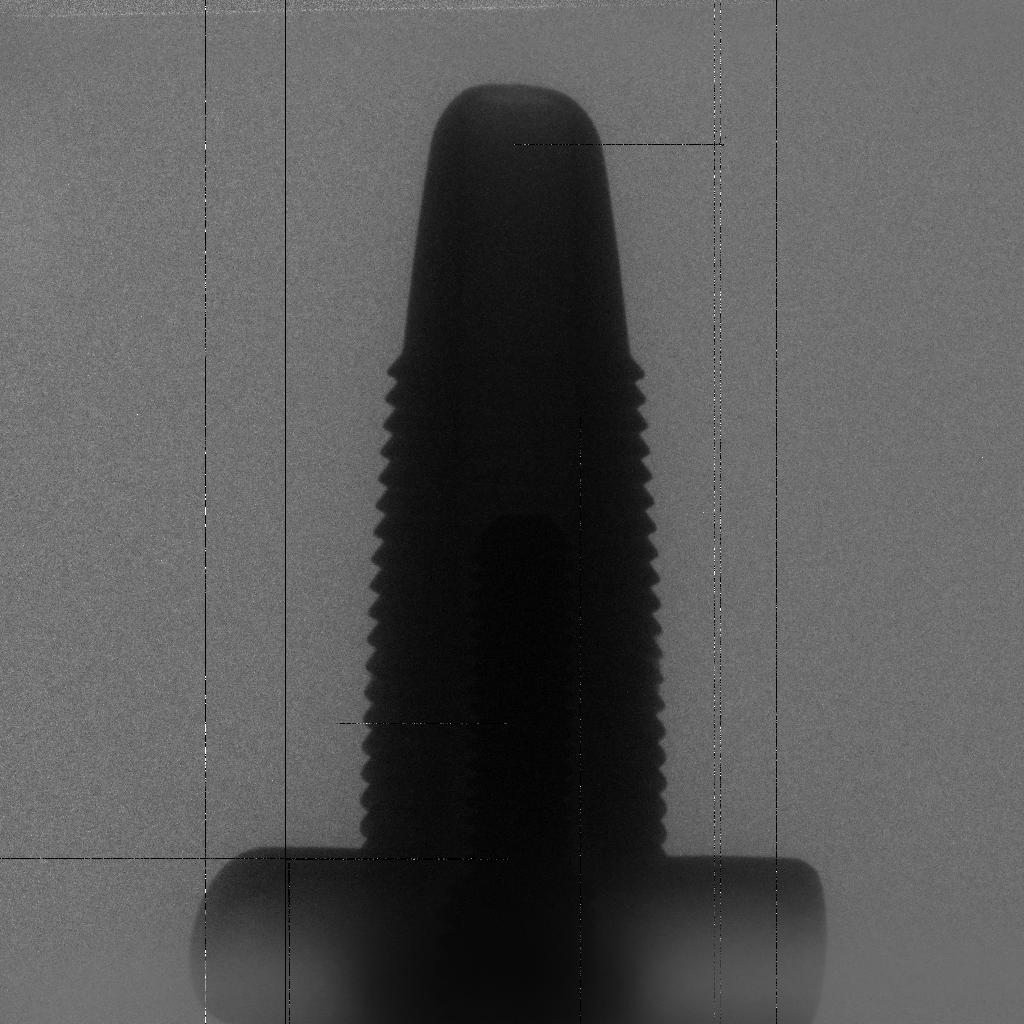}
\end{center}
  \caption{X-Ray Absorption Image taken with the X-ray source. The image is obtained after irradiating some 30 seconds at 1 Khz.}
  \label{Radiography}
\end{figure}
In Figure \ref{Radiography}, we can observe an absorption image performed with the presented X-ray source and a X-ray CCD array detector (RadEye 2, low dose:10-50 keV). Besides, the source stability achieved is good enough to perform also tomographic experiments. Moreover, the X-ray source is a pulsed source with very short duration, so it is possible to perform pump-probe experiments with infrared and X-rays simultaneously. It is clear that the field of application of these novel sources is growing up.

\section{Conclusions}
In this work, we have presented the new X-ray laser driven source recently developed at the L2A2. The source is based on a back-scattered X-ray emission of a hot over dense plasma induced by 1 mJ, 35 fs laser pulse on a Cu solid target. To guarantee material refreshment at 1 kHz repetition rate,  the target is mounted on a fast rotatory positioning system. The target has to be placed on the laser focus during the movement, and in order to avoid the wobbling we have presented  solutions for target mounting. We have implemented a speckle technique mixed with a best focus determination based on X-ray spectra, which allows us to take into account the effect of the laser self-focusing in air. We have also demonstrated a way to increase the source stability, operating with a variable rotation speed. A complete characterization of a X-ray spectrum is also presented, showing the characteristic Cu K peaks and the Bremsstrahlung continuum composed by two different temperatures ($T_{cold}$=17.7 keV and $T_{hot}$=29.3 keV) as it is characteristic in this kind of laser-plasma interaction. To finish, we have also shown the first application of our X-ray source with a X-ray absorption image. In future, we plan to obtain tomography and phase contrast images with the X-ray source presented here. However, the source presented here is not only interesting for imaging purposes, also for going deeper in the physics of the plasma, the different distributions of electrons and the interactions among the laser and the target.

\ack
This work has been supported by Spanish Ministry of Industry within the framework of the RETOS project (RTC-2015-3278-1) and by Xunta de Galicia and EU, LASERPET project (2013-AD009.01). Camilo Ruiz acknowledges the MINECO project FIS2016-75652-P.

\section*{References}
\bibliography{biblioLambdaCube,RayosX_targ3}
\bibliographystyle{iopart-num}
\end{document}